# Agent based simulations visualize Adam Smith's invisible hand by solving Friedrich Hayek's Economic Calculus.

Klaus Jaffé

Universidad Simón Bolívar

Caracas, Venezuela

kjaffe@usb.ve

**Abstract:** Inspired by Adam Smith and Friedrich Hayek, many economists have postulated the existence of invisible forces that drive economic markets. These market forces interact in complex ways making it difficult to visualize or understand the interactions in every detail. Here I show how these forces can transcend a zero-sum game and become a win-win business interaction, thanks to emergent social synergies triggered by division of labor. Computer simulations with the model Sociodynamica show here the detailed dynamics underlying this phenomenon in a simple virtual economy. In these simulations, independent agents act in an economy exploiting and trading two different goods in a heterogeneous environment. All and each of the various forces and individuals were tracked continuously, allowing to unveil a synergistic effect on economic output produced by the division of labor between agents. Running simulations in a homogeneous environment, for example, eliminated all benefits of division of labor. The simulations showed that the synergies unleashed by division of labor arise if: Economies work in a heterogeneous environment; agents engage in complementary activities whose optimization processes diverge; agents have means to synchronize their activities. This insight, although trivial if viewed *a posteriori*, improve our understanding of the source and nature of synergies in real economic markets and might render economic and natural sciences more consilient.

# I. Introduction

Adam Smith in his book The Wealth of Nations (Smith 1776) described the operation of the market as follows: "*Every individual necessarily labours to render the annual revenue of the society as great as he can. ... He, however, does not have the slightest intention of promoting the public interest or is aware that he is promoting it. He intends only his own gain and is led, as in many other cases, by an invisible hand that makes him promote a cause that does not form part of his intentions. This is not a disadvantage for society. By pursuing his own interest, he frequently promotes that of the society more efficiently than if his interest were the latter. I do not know of much good dispensed by those who strive to represent the common good. It is not from the benevolence of the butcher, the brewer, or the baker, that we can aspire to our dinner, but from their attention to their own interests*". In another part of the book he writes "*The greatest improvement in the productive powers of labor, and the greater part of the skill, dexterity, and judgment with which it is anywhere directed or applied, seem to have been the effects of the division of labor .... It is the great multiplication of the productions of all the different arts, in consequence of the division of labor, which occasions, in a well governed society, that universal opulence which extends itself to the lowest ranks of the people*" These are beautiful descriptions of phenomena where the interactions at the individual level bring as a consequence dynamics significant only at the social level, without individual activity being conscious of it. It is a fascinating phenomenon but difficult to study using traditional experimental techniques.

The discovery of *the invisible hand of the market* is a major achievement of humankind. It recognizes the absence of centralized social cohesive forces and discovers forces of the market that explain our social dynamics. More detailed studies of the effects of division of labor have been published (Becker and Murphy 1994, for example), but all failed to grasp analytically in ways acceptable to the natural sciences, the emergence of synergies in economic markets due to division of labor. This inability to grasp numerically these phenomena has led some economists to conclude about the analytical intractability of all details in complex economies. Prominent among these thinkers is Friedrich Hayek (1948), who coined the term "Economic Calculus" when referring to this fundamental analytical limitation of economic analysis. He said (Hayek 1961) "*[economics] has become too ambitious by applying standards of rigorousness ... to the empirical science of*

*economics where there are definite limits to what we can positively now; that we shall see more clearly what economics can do if we separate that logical groundwork – the economic calculus as I have called it – from its use in the empirical science of economics; and that, though this science is of great help in all-important issues of the choice of an economic order and of the general principles of economic policy, its power of specific prediction is inevitably limited – limited by the practical impossibility of ascertaining all the data – those very data whose utilization in the allocation of resources is the great merit of the market system*".

Beyond economics, the effect of the behavior of the individual on the performance of the social aggregate and how these interactions might led to the emergence of novel properties, can be studied from the point of view of Complex System Science (Jaffe 2014). Specifically "artificial societies" or computer simulations of social dynamics (Tesfatsion 2006, Magliocca et al. 2014 for example) have shown their worth in illuminating how the aggregate of various simple interactions might produce phase transitions and the emergence of novel properties of the system and even novel phenomena. These modern computer simulations, specifically agent-based simulations, allow us to explore complex economic phenomena. Examples include the complexity of exchanges (Axtell 2005), and money dynamics and banking catastrophes (Brummitt 2014). However, agent based simulations have not been incorporated in mainstream economics (Leombruni and Richiardi 2005), nor have they unveiled until now in detail the working of the invisible hand of the markets. Agent based simulations are a powerful tool in clarifying fundamental aspects of the working of complex economic phenomena (Jaffe 2014). Its potential in visualizing fundamental concept in very simple economies will be explored here so as to avoid the limits "*of ascertaining all the data whose utilization in the allocation of resources is the great merit of the market system*" (Hayek 1911), and which are not easy to determine analytically in more complex economic settings.

Simulations with Sociodynamica allow for exploring abstract virtual economies that are far simpler than real ones but already so complex that the experimenter may looses the integral view over the interactions between environment, agent behavior, pricing mechanisms, and environmental heterogeneity in the market. This might happen in simulations of economies were realistic price dynamics were included that showed that division of labor was the strongest predictor of successful economic performance (Jaffe 2015). However, it was not clear in these simulations, if this effect of division of labor was exerted through the price dynamics of the economy or through other means. In order to pinpoint the source of the synergies achieved by division of labor, the model was simplified until the emergent effect of division of labor disappeared. Stripping out effect of pricing on the market dynamics did not eliminate the effect of division of labor. This allowed us to follow in detail the features that make division of labor work, making the system amenable to analytical analysis, solving the required "Economic Calculus" (Vaughn 1980) for this simple virtual economy.

**2. The Model**

The agent based computer simulation model Sociodynamica is a freely available agent based simulation model written in Visual Basic that has previously been used to study the effect of altruism and altruistic punishment on aggregate wealth accumulation in artificial societies (Jaffe 2002a, 2004a, 2008, Jaffe and Zaballa 2009, 2010) and to grasp the dynamics of complex markets (Jaffe 2002b, 2004b). These models are completely mechanical in nature, and individual incentives may emerge trough an evolutionary process that makes agents with the right combination of incentives or behaviors survive, and those with the wrong combination, to become eventually extinct. The features revealed by Sociodynamica are very similar, and in cases identical, to those revealed by Sugarscape, an agent based model developed independently by Axelrod (1997); or the model developed by Axtell (2005). In all three cases, Walrasian solutions in which an auctioneer centrally computes prices cannot be made more efficient than the decentralized alternatives based on free and heterogeneous agents making these decisions. These results supports the proposition of Adam Smith's that markets are ruled by an Invisible Hand that coordinates the different kind of labor rendering markets efficient. Specifically, simulations in complex economic setting showed that omnipotent agents performing all tasks, produced less aggregate wealth than simulations where three different agents performed different tasks, such as farming, mining and trading (Jaffe 2015). This counter-intuitive result was partly due to the fact that optimal prices and conditions for trade were different for each agent, depending on its spatial position in the virtual world. Omnipotent agents had to assume average solutions to balance their different tasks. Therefore, they never traded at optimal prices and optimal quantities according to their spatial position. Here the model was simplified, until the effects of prices dissipated, to reveal fundamental economic features that allow the emergence of synergies from division of labor.

The model simulates a virtual society of agents who farm and mine for foods and minerals respectively, analogous to the model "Sugarscape" by Axelrod (1984), and also trade their surplus according to different economic settings. The agents inhabit in a continuous flat two-dimensional toroidal world (see Figure 1) that was supplied with patches of agricultural land ("sugar" or "food") and separate non-overlapping patches of mines ("spices" or minerals). Diverse agents were distributed at random on a fine-grained virtual landscape with resources. Simulations depended on the type of movement of agents, and thus, to simplify interpretation of results, agents were simulated as immobile entities. Their individual utility function was defined by two resources. Each time step, any agent that happened to be located over one of these resources, acquired a unit of the corresponding resource, accumulating its wealth, either as sugar or food ($G_1$) and/or as spices or minerals ($G_2$). Agents spend a fixed amount of each resource in order to survive, consuming each of them at a basal constant rate (default value was set to 0.1 units of the corresponding resource at each time step). Both resources were consumed and metabolized similarly, but food was 3 times more abundant than minerals (the size of the patch for minerals was set to 100 x 100 pixels and for food was 300 x 300 pixels). Each patch remained in the same place during each simulation run and the resources inside them were replenished continuously. Agents perished when they exhausted any of the two resources. Success in gathering and trading resources was defined by variables that produced behaviors that made them unable to compete successfully for resources. These variables included type of movement, spatial positioning, price thresholds for selling each of the resources, price threshold for buying the resources, and type of agent. During the simulation, natural selection weeded out unsuccessful combinations of these variables. The total population of agents was maintained constant by creating the required amount of new agent necessary, each with randomly assigned initial parameters. Initial parameters were the type of agent, the random spatial position and the initial amount of money used start trading resources (the default initial value was set to 10 units of money). The amount of money for each agent varied according to its trade balances. Agents gain money when selling food and/or minerals and lose money when buying them.

Agents traded the resources they possessed with other agents. In order to trade, they had to find a partner with the desired resource, and they had to have agreement over prices. The trade could be among any agent in a population of omnipotent agents without "division of labor". When simulating division of labor, agents specialized in collecting food or collecting minerals, or collecting neither but engaging only in trade. Here, agents were subdivided into three categories. Farmers which specialized in collecting only resource 1; Miners which extracted only resource 2; Traders specialized in trading minerals for food when encountering a farmer, and food for minerals when encountering a miner. Food collectors traded only with mineral collectors and traders, mineral collectors traded only with food collectors and traders, and traders could interchange resources with all types of agents. Trades were allowed only between agents spaced at a distance not larger than the "contact horizon" of the trading agent. Each time step, all buyers searched for potential sellers of the required good by contacting randomly up to 10 agents in the area defined by this contact horizon. If finding a seller with the wanted goods at or below the price defined by the buyer, a trade was executed using the price of the seller. Trades were limited to the amount of money available to the buyer and the amount of goods possessed by the seller, unless credit was simulated. Variation of this contact horizon allowed to simulated different levels of globalization or integration of economic agents. The effect of the degree of globalization (or the size of the market) on the economy can thus be measured quantitatively, a feature that is not possible with real economies (but see Campos et al. 2014).

Prices were initially assigned to each agent for each resource at random from a range of values defined by the experimenter, and then varied according to supply and demand as experienced by each individual agent. That is, at the end of every time step, after finishing a tournament of trades in the market, each selling agent attempting to sell parts of its excess of resource that could not find a willing buyer because of the price it asked for, reduced its reference price by an unit. And each buyer that could not find a seller willing to sell the desired resource at the desired price increased its reference price for that resource. In this way, each agent maximized its self interest by selling each resource at the maximum price possible and buying at the lowest.

Various processes were simulated. A first process of the simulation was the balance between income (I) of resources (r) and their consumption (C). For survival, agents were required to conform to

$$I_r > C_r \qquad (1)$$

Income can be either by direct gathering (G) or by trade (T)

$$Ir = Gr + Tr \qquad (2)$$

Here each agent has to balance two resources in order to survive. I simulated a utility function (U) so that U had to remain positive for the agent's survival and:

$$Ur = (Gr + Tr) - Cr \qquad (3)$$

A second process of the simulation was the dynamics of traded resources. These resources can increase or decrease, according to the balance of resources bought (B) and sold (S)

$$Tr = Br - Sr \qquad (4)$$

The amount of resources bought and sold depends on the availability of money (M) and the price (P) paid for the resource by the agent (a)

$$Br = M_a / Pr_a \qquad (5)$$
$$Sr = M_a / Pr_a \qquad (6)$$

The amount of money of agent (a) depends on the amount spend buying (Mbr) and the amount gained selling (Msr) resources

$$M_a = M_a 0 + Msr - Mbr \qquad (7)$$

where $M_{a0}$ stands for the initial amount of money supplied to each new agent j

Ur was calculated every time step for the population of agents (a) so that the total accumulated of wealth for each resource (Wr)

$$Wr = \Sigma_a Ur \qquad (8)$$

Agents with $U_1 <= 0$ or $U_2 <= 0$ were eliminated and substituted by new ones with default properties, as an analogy of broken companies that are replaced by new start-ups.

Here we focus on the age of the agents as the most relevant variable for assessing the benevolence of an economic system. The average age is a measure of the probability of survival of individuals in the population, but other measures are possible (Jaffe 2015). The aim was to pinpoint the features that allowed the emergence of the synergies of the market due to division of labor. Simulations of virtual worlds with one type of agents (Omnipotent agents), two types (Farmers + Miners), and 3 types (Farmers + Miners + Traders) allowed determining the effects of increasing complexity of labor structure. Simulation with homogeneous and heterogeneous distribution of resources allowed to asses the effect of the environmental complexity, and simulations with different contact radius provided insights to the importance of synchronization between trading agents.

A longer and more detailed description of the simulations is provided in Jaffe (2015) and the detailed program in Visual Basic is available in the help feature of the program. Simulations can be run with parameters choose at will by downloading Sociodynamica at [http://atta.labb.usb.ve/Klaus/Programas.htm]. Default values used here were: Contact horizon for transaction = 200. For both resources, Initial prices = 3 units, Reserve units not traded = 1. Amount of resources metabolized per time step = 0.1, Amount of resources collected per time by agent = 2. Number of agents = 500, maximum number of trades per time step = 10. More details are given in the appendix. Videos of the simulations run for the present paper are available at:
http://atta.labb.usb.ve/Klaus/EC/ECVideos.html
A user friendly game version of Sociodynamica is available at
http://www.bcv.cee.usb.ve/juegos/intro_en.html

3. Results

Figure 1 shows two examples of the output of simulations with free prices after 200 time steps. The upper figure shows the virtual world when simulating omnipotent agents; the second figure reflects the outcome when division of labor (Farmers + Miners) was included in the simulations. The figures reflect the effects on the economic dynamics of introducing division of labor. In the figure with omnipotent agents, agents over mineral fields are smaller (have less wealth) than those over fields with food. In contrast, the figure from simulations including division of labor showed that miners over field of minerals were very wealthy, and so where many farmers over field with food. The price agents were willing to pay for minerals was higher among farmers (red borders in Figure 1B) and lower among miners (blue borders in Figure 1B) when division of labor was simulated. In the case of the omnipotent agents, prices agents were willing to pay for minerals or food seemed to be randomly distributed (mix of colors in borders of agents in Figure 1A). This example reveals that omnipotent agents made sub-optimal trading decisions. They sold the resource they had accumulated in more abundance to any other agent willing to take it; whereas specialized agents (Figure 1B) traded only the resource they were collecting; that is, Miners only sold minerals to Farmers and Farmers sold only food to Miners. The trading patterns of omnipotent agents produced less wealth in the long term that that of specialized agents, even though the cognitive complexity of the algorithm omnipotent agents used was more complex.

**Figure 1**: Representation of the landscape of two virtual economies. Figure 1A shows a result from a simulation with free prices in an economy collecting food and minerals by omnipotent agents; whereas Figure 1B shows the same but for agents specialized either in mining or farming.

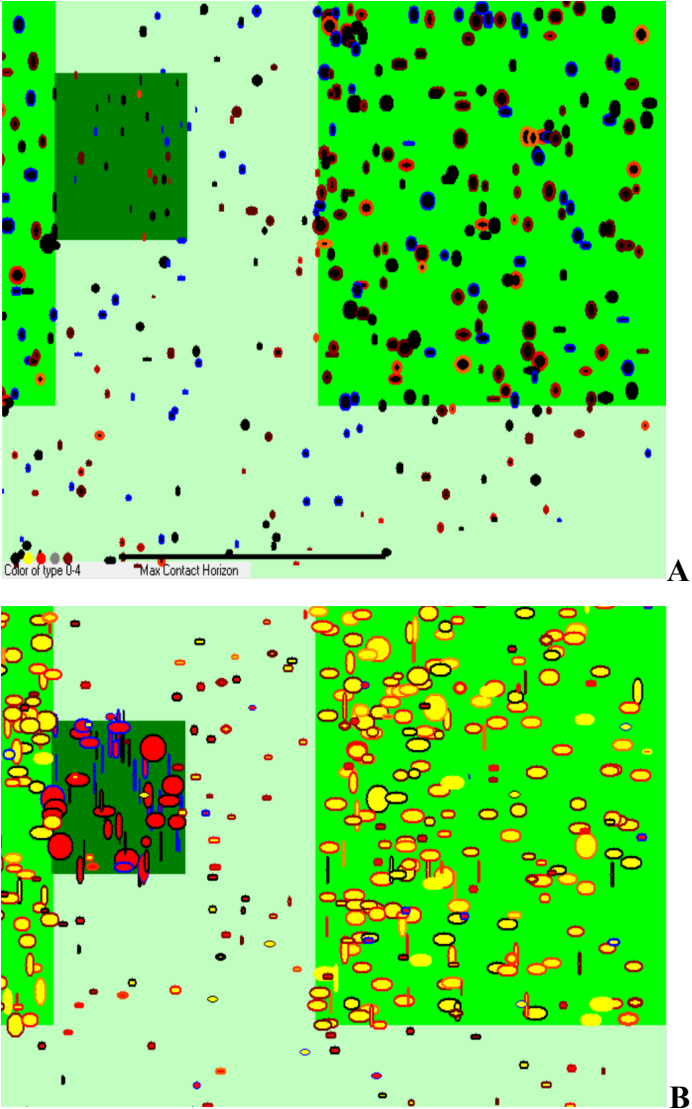

- Bright green field is covered with "Food"; darker green field is covered with "Minerals"; the lightest green is devoid of resources.
- Each agent is depicted as a colored sphere. The color of the body of the sphere describes the type of agent: Farmers are yellow, Miners are red, and Omnipotent agents are black. The width of the bubble is proportional to the amount of food and minerals accumulated by the agent and the height by the amount of money the agent possesses.
- The thickness of the border of the sphere is proportional to the perceived cost of living calculated as Food Price + Mineral Price and the color of the border rage from blue to yellow. The more reddish or even yellow the higher the ratio between the Food price and Mineral price. Agents with red and yellow borders pay more for minerals, whereas those with blue or black borders pay less for minerals compared to what they are willing to pay for food.

**Table 1**: Average age accumulated by agents after 200 time steps in simulations exploring the effect of fixed or free prices, with omnipotent agents or with agent dividing labor. Maximum contact radius of agents was 200 pixels. Each data is the mean of 100 simulation runs.

|  | **Omnipotent agents** | **Division of labor 2 Farmers + Miners** | **Division of labor 3 Farmers+Miners+Traders** |
| --- | --- | --- | --- |
| **Fixed Prices** | 11.2 ± 0.71 | 86.9 ± 3.5 | 69.2 ± 3.6 |
| **Free Prices** | 11.3 ± 0.8 | 87.7 ± 3.2 | 69.1 ± 3.9 |

The difference between the two simulations shown in Figure 1 is presented quantitatively in Table 1. As intended in this simplified model, the effect of centralized or decentralized pricing was stripped out, so that this feature did not affect results. Agents in simulations with 2 type of agents (Farmers and Miners) performed much better economically, achieving much longer life-spans in average that agents in simulations with only one type of agents (omnipotent agents). Increasing the complexity of the division of labor by including Traders - i.e. agent that did not collect resources but only bought and sold them, using the same algorithm as that of omnipotent agents - decreased the efficiency of the resulting economy as reflected by lower average life-spans of agents.

Eliminating heterogeneity in the environment eliminated the advantage of division of labor (Table 2). Simulations where fields of minerals overlapped in all its extension with field of food, produced quantitative results that were identical between virtual economies of omnipotent agents and economies with division of labor with two type of agents ($p > 0.98$ for rejection of null hypothesis using Student's t-test). Introducing Traders in addition to Farmers and Miners reduced the economic efficiency of division of labor somewhat.

**Table 2:** Average age accumulated by agents after 200 time steps in simulations exploring the effect of homogeneous economic environments. Maximum contact radius of agents was 200 pixels. Each data is the mean of 100 simulation runs.

| **Omnipotent agents** | **Division of labor 2 Farmer+Miner** | **Division of labor 3 Farmer+Miner+Trader** |
|---|---|---|
| 197 ± 4 | 198 ± 4 | 176 ± 7 |

Introducing the possibility of different motility among agents also affected the difference between between simulations of only omnipotent agents and simulations where division of labor occurred: Selection tended to favor mobile traders and immobile farmers and miners. Results suggested that the division of labor, in order to work properly, required an adequate coordination of actions or synergy between the agents. For example, omnipotent agents sold whatever resource they had, independently of what they collected, whereas farmers and miners only sold the product they collected and bought the one they did not collected. Also, traders in simulations with no mobile agents seemed just to introduce noise in the economy as they were not allowed to behave differently from miners and farmers. In Figure 2 the effect of an improved ability to trade is shown. Here, the maximum contact radius determining the distance at which potential traders could be spatially separated was varied. As expected, results show that at greater maximum contact radius, the economy performed better. This trend, however, was not linear. Prices had a very strong non-linear relationship with the contact radius. This relationship is strongly depended on the topology of the resource distribution simulated, as was shown before (Jaffe 2015).

**Figure 2**: Effect of the contact radius on the economic performance of the virtual economy, measured by the average age (blue circles, left scale) and average price of food red squares, right

scale) accumulated by agents after 200 time steps in simulations with free prices and with division of labor.

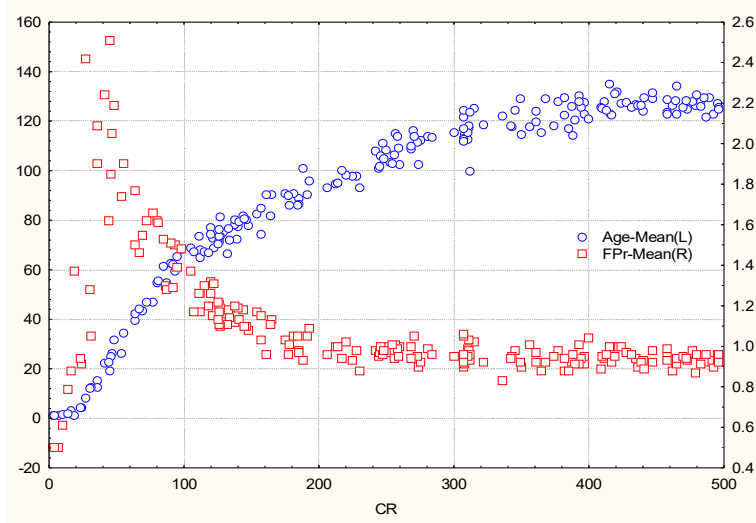

## 4. Conclusions

The results presented here showed how this relatively simple simulation model of a virtual economy, identified features that are indispensable for making the emergence of synergies due to division of labor possible. The most important Conditions for synergy to arise are:

1- Spatial or temporal heterogeneous environment and/or behavior
2- Complementary activities with divergent optimization
3- Synchronization of one or more heterogeneities

The first feature is interesting in that it recognizes that the complexity of labor structure mirrors the complexity of the economic environment where it works, and *vice versa*. This feature is a consequence of economic activity being molded by the environment to which it has to adapt. In human societies economic wealth of a country and economic complexity are linked (Hausmann and Hidalgo 2014). Economic complexity, of course, can only be achieved with complex division of labor.

The second feature is illuminating as it allows predicting when division of labor might improve economic activity and when not. Division of labor might prevent inefficient trades and/or might make more efficient trades possible. Tasks, which converge in skills, might not require specializations, whereas tasks that require very different types of skills will benefit more of division of labor. It can be argued that more intelligent agents might be capable of performing several tasks and thus, rather that division of labor, economies benefit from more complex or intelligent agents. This might be true but complexity and intelligence have their costs and might be sub-optimal if different simple agents can handle the problems with higher efficiency. This seems to be the case in the social evolution of ants, where a negative correlation between brain development of individuals and social complexity was evidences (Jaffe and Perez 1989).

The third feature is possibly the more difficult to manage in real situations. Experience shows that on-time synchronized productive chains allow specialization of tasks to be more productive overall; whereas low division of labor is more tolerant to inefficient supply chains. These features explain many a difference between highly developed economies and ones with incipient industrialization and poor services.

The identification of these 3 features as fundamental in allowing division of labor to elicit economic synergies might seem trivial. But browsing the literature, a great number of reasons have been postulated to explain the synergies created by economic markets. The problem seems far from solved (see Kochugovindan & Vriend 1998), Even if found to be trivial *a posteriori*, the simulations helped to identify the relevant features and discard superfluous ones. Bowles (2009) for example, defined the invisible marker mechanism as a Nash equilibrium and its Pareto optimal, where the self interest of each actor yields an outcome that maximizes the well being of each. Here we reveal in more detail how this can be achieved. The three features unveiled here seem to be very general, relevant to system dynamics, biology, human society, and real economies. Possible empirically falsifiable predictions based on these tree features are:

1- Division of labor should be more developed in societies that exploit a greater variety of resources.

2- Better communication and economic instruments are provided by more sophisticated financial systems which in turn provide better opportunities for synergies between economic actors in more complex economies. Novel communication technology, such as the Internet, can also broaden the contact radius of economic agents improving synchrony.

3- More division of labor leads to more incompatibilities between skills required to perform them and thus to more diverse specialized education.

4- Larger contact horizons or more globalization improves the synergies unleashed by division of labor as they broaden the scopae for more diverse interactions between economic agents

Empirical evidence found so far would support these predictions (examples for each of the three point are respectively: Hausmann and Hidalgo 2014, Mantegna and Stanley 2000, Dale 2005). Though research purposefully designed to answer these questions should be designed. We know that division of labor is related to synergies associated with fundamental aspects of social systems in biology: Thermodynamic studies showed that the amount of entropy of ant societies and the levels of division of labor are related (Jaffe and Heblin-Beraldo 1993). In human society, economic synergy and division of labor have important relationships. Research showed that more complex economies requiring more division of labor accumulate more wealth and produce higher economic growth (Hausmann and Hidalgo 2014). Empirical data showed that an even better predictor than economic complexity for future economic growth in developing countries is the scientific knowledge estimated by the amount of academic activity (Jaffe, Rios, and Florez 2013). Even more striking, the type of division of intellectual activity in a country is a much better predictor than the total complexity or the absolute amount of academic research performed (Jaffe et al. 2013). That is, division of academic labor that prioritize basic natural sciences over applied sciences and social sciences, is much more efficient in producing future economic growth. These results, in the light of the findings of the present simulations, show that much remains to be learned about the quality and quantity of division of labor and its effect on economic activity. More interdisciplinary research is needed to improve our understanding of this very fundamental phenomenon.

This exercise shows that computer simulation of simple economic agents can generate a non-linear dynamics that resembles real life features of known economic system. Simulations of very complex systems produce complex results that may become intractable even with sophisticated statistical analysis. Here we overcame this limitation by focusing on very specific and fundamental problems. The simulations presented revealed fundamental features that allow division of labor to create economic synergies. This insight was possible by solving some aspects of the "Economic Calculus" (Hayek 1991) in a very simple system. This feat is impossible in complex real situations, but the insights gained in simple systems help in understanding synergies in more complex ones. Simulation models, besides having a potential in experimental economic research, are a fantastic tool to make complex phenomena visible to human understanding and thus should have a potential, if properly adapted for that purpose, in didactic games for teaching economics at all levels of educational and academic specialization. Science learned through games based on simulations might reduce self-serving cognitive biases among lay people, professional practitioners and decision makers, improving the rationality of our society and thus, hopefully, it's economic performance.

**Acknowledgments:** Thanks are due to Stephen Davies and Juan Carlos Correa for helpful comments on previous versions of the manuscript

# Appendix

**Sociodynamica** creates a virtual society where **a**gents exploit and compete for resources and share resource 1 among them, according to the settings defined by the [internal parameters](internal parameters) and the [external parameters](external parameters). The agents may acquire renewable and non-renewable resources trough work; they may accumulate those resources and commercialize them. At the same time, agent may acquire resources through commerce.

## *Global Parameters*

**POP:** Number of agents (no)

**TAR:** Aggregate total wealth accumulated by all agents

## *Simulation logic*

**Each time step:**
Do simulation loop
Matrix:
    Eliminate variable types previously defined
    Eliminate agents with wealth = 0 (par 11)
    Increment age agent(i, 7) = agent(i, 7) + 1
    Use of resource 1 and 2 agent(i,11)- BRC1; agent(i,12) – BRC2
    Assessment of GDP GDP = GDP + agent(i, 11)
    Show and Plot

*Plot* Shows the agents according to their total resources (Food+Money) indicated as the sqr of the diameter.
The high of the agent is proportional to the total wealth of the agent's money.
The color of the bubble depends of the type of agent as indicated at the left bottom of the screen
The thickness of the border is proportional to the perceived cost of living (Prize for food + Price for minerals).
The color of the border is more reddish or even yellow the higher the ratio MinPrice/FoodPrice.
Blue borders indicate that food prices are higher than mineral prices.
Black bar at the bottom indicates a length of 100 pixel

## **Internal parameters:**

*General parameters* (number in parenthesis indicates the column in the master matrix)

| | | |
|---|---|---|
| X (0) | Spatial dimension1 | |
| Y (1) | Spatial dimension2 | |
| CRa (2) | **Contact Radius or Contact Horizon** | Maximum distance at which interchange between agents example) |
| TMo (4) | **Type of Movement** | Type of spatial displacement: NO MOVEMENT |

*Characteristics of agents*

| | | |
|---|---|---|
| Age (7) | **Age** | Age of agent |
| WT (10) | **Wealth-Money** | Total capital in liquid money |
| WFo (11) | **Wealth-Food** | Amount of resource 1 (Renewable). If WFo(i) = 0 then agent i starves (i is eliminated) |
| WCo(12) | **Wealth-Commodity** | Amount of resource 2 (Minerals or Non-Renewable resource) |
| Dept (13) | **Dept** | Accumulated Dept |
| Well (19) | **Well-being** | Amount of resources 3 = r1 * r2 * Gain |
| TAg (20) | **Type of Agent** | Specialization or task of agent |

    0 Omnipotent: exploits resources 1 and 2
    1 Farmer: exploits only resources 1
    2 Miners: exploits only resources 2
    3 Trader: does not collect but trades and provides credit: i.e creates money

**External Parameters**

*General*

**Initial Nr. of Agents** (ino)

**Optimum Population Size** (ops): Maximum number of agents aimed at through ssconst

**Simulation Scenario**: production of new agents (ssconst):
    0: New agents are created each time steps, until the number indicated by ops is achieved. New agents are assigned internal parameters at random

**Proportion Culled** (PC). The proportion of agents killed randomly when population in excess of ops

**Dangers**, other than starvation danger. Large values increase random selection; large WCo reduces this.

**Fitness function**: Agents, in order to continue in the virtual word, had to satisfy each time step the rule: 100 * Rnd / (amount of resource 2) / (mean wealth of resource 2) < 1000 * Rnd /

**Dangers:** probability of being eliminated in random selection events

*Resources 1 and 2*

**Resource 1 (provides WFo) and 2 (provides WCo)**
**Number of patches of Resource (**RNR) Number of resource patches
**Size of patch of Resource** (SNR) Maximum sizes of each patch (but see Mutation)
**Degradation of Resource** due to consumption (DNR) Amount lost due to consumption (RD)
**Distribution Pattern of Resource** (DPR) Resource is distributed:
  1: Fixed size, randomly distributed
  2: Fixed size, centered
  Else Random size, randomly distributed
**Basal rate of metabolism** (BRC) Amount of resource passively used-up (b)
**Efficiency of consumption** (EfC) Amount of resource assimilated.
   When EfC>10 then productivity is simulated: Agent collects = (EfC-10)*price (either for resource 1 and/or 2)
**Frequency of change in distribution (**FCh) Frequency in t-steps distribution changes
**Consumption of resource**: Rate of exploitation: resource = resource – DNR
   BRC: Wearing or passive use of resource agent- BRC
   Resource 1 can be modeled as a renewable resource (agriculture for example),
   whereas resource 2 as a non renewable resource (mining for example),
   by assigning DNR = 0 and DNR = 1 respectively

**Food Reserve** (FR): Minimum amount of food needed for agent to engage in transactions of any kind

**Min Food for Reproduction** (MFR): The amount of food that need's to be accumulated before reproduction can start when Simulation Scenario is 1 or 4

**Type of economy (EconoT)**
  **0**: No Barter nor any other interactions except taxes
  **1**: **Barter**: with no money
  **> 1**: Money as Species. If **Price adjust** = 0 then Fixed prices
  If **Price Adjust** > 0 the Prices are determined by demand: agents not selling decrease price by one unit; agents not finding seller from which to buy increase price by one unit.
  **3**: + Financial: Traders lend money.
  **4**: + Agents finishing a successful sale, increase price by one unit
  **5**: + Successful buyers decrease their future asking price by one unit
  **6**: + As in EconoT 4 and 5
  **7**: + As in 2, 3, 4 and 5

8: As in 2 and 3

**10:** Agents pay Taxes: Taxes collected are increased synergistically by Taxpool * SSTax prior to their distribution.

**Food price**: in integer units

**Mineral price**: Price of commodities in integer units

**Price Adjustment**: Units prices are reduced or increase when trade fails. 0 = fixed prices